\documentstyle[twocolumn,seceq,epsf]{jpsj}

\def\runtitle{Effects of In-Plane Impurity Substitution in Sr$_{2}$RuO$_{4}$}
\def\runauthor{Naoki {\sc Kikugawa}, Andrew Peter {\sc Mackenzie} and Yoshiteru {\sc Maeno}}

\title
{
Effects of In-Plane Impurity Substitution in Sr$_{2}$RuO$_{4}$
 }

\author
{ 
Naoki {\sc Kikugawa}$^{1,2}$, Andrew Peter {\sc Mackenzie}$^{3}$ and Yoshiteru {\sc Maeno}$^{2,4}$
}

\inst
{
$^{1}$Venture Business Laboratory, Kyoto University, Kyoto 606-8501\\
$^{2}$Department of Physics, Kyoto University, Kyoto 606-8502\\
$^{3}$School of Physics and Astronomy, University of St. Andrews, St. Andrews, Fife KY16 9SS, Scotland\\
$^{4}$International Innovation Center, Kyoto University, Kyoto 606-8501\\
}

\recdate
{
\today
}

\abst
{
We report comparative substitution effects of $nonmagnetic$ Ti$^{4+}$ and 
$magnetic$ Ir$^{4+}$ impurities for Ru$^{4+}$ 
in the spin-triplet superconductor Sr$_{2}$RuO$_{4}$. 
We found that both impurities suppress the superconductivity completely 
at a concentration of approximately 0.15\%, 
reflecting the high sensitivity to translational symmetry breaking 
in Sr$_{2}$RuO$_{4}$. 
In addition, 
a rapid enhancement of residual resistivity is in 
quantitative agreement with unitarity-limit scattering. 
Our result suggests that both nonmagnetic and magnetic impurities in Sr$_{2}$RuO$_{4}$ act as 
strong potential scatterers, similar to the nonmagnetic Zn$^{2+}$ impurity 
in the high-$T_{\rm c}$ cuprates.  
}
\kword
{
Sr$_{2}$RuO$_{4}$, impurity effects, unitarity scattering 
}

\begin{document}
\sloppy 
\maketitle

%\clearpage

%\baselineskip=2\baselineskip
%%%%%%%%%%%%%%%%%%%%%%%%%%%%%%%%%%%%%
%
%Impurity Effect in Unconventional Superconductivity 
%
Unconventional superconductors such as heavy fermion compounds and 
high-$T_{\rm c}$ cuprates have attracted much attention in the last two decades,  
since highly correlated $f$- and $d$-electrons in these materials 
play essential roles in the emergence of the unconventional superconductivity \cite{Mathur}. 
In contrast to conventional superconductivity with $s$-wave symmetry, 
nonmagnetic impurities as well as magnetic impurities 
act as strong pair breakers and severely suppress the transition temperature $T_{\rm c}$ 
of unconventional superconductivity. 
The suppression of $T_{\rm c}$ reflects the sensitivity to 
translational symmetry breaking \cite{unconventionalSC} 
and is a characteristic of anisotropic pairing. 
%
%Moreover, 
%systematic study of the substitution effect by impurities for unconventional superconductors 
%has been recognized as a powerful method in order to 
%reveal prominent electronic states reflecting the strong correlations 
%by host $f$- \cite{UPt3} and $d$-electrons \cite{high-Tc}. 
%
%
Systematic studies of impurity substitution have also been used to obtain information 
on the underlying strongly correlated electronic states 
in both $f$- \cite{UPt3,HFSC} and $d$-electron systems \cite{high-Tc}. 
%

%
%Sr2RuO4
%
Here, we study the effects of such substitution in the layered perovskite ruthenate Sr$_{2}$RuO$_{4}$, 
whose superconductivity \cite{Nature} is unconventional, 
most probably involving spin-triplet pairing \cite{physicstoday}.  
%
%The superconductivity of layered perovskite ruthenate Sr$_{2}$RuO$_{4}$ 
%with intrinsic transition temperature $T_{\rm c0}$ = 1.5 K \cite{Nature} 
%is identified as unconventional pairing symmetry, 
%most probably spin triplet \cite{physicstoday}, 
%in contrast to $d$-wave symmetry in the cuprates with the same crystal structure. 
%
The normal state properties of Sr$_{2}$RuO$_{4}$ are described quantitatively within 
the framework of a quasi-two-dimensional Fermi liquid \cite{Maeno} 
with the Fermi surface consisting of three nearly cylindrical sheets 
($\alpha$, $\beta$ and $\gamma$) \cite{Andy-dHvA}. 
Comparison with the band-structure calculation \cite{Band} indicates that 
strong correlations among the electrons originating from the Ru$^{4+}$ ions 
(4$d^{4}$ in the low spin configuration) hybridized with $p$-electrons of 
surrounding oxygen play an essential role in  
the physical properties of Sr$_{2}$RuO$_{4}$. 
%

%
%Past Study of the Impurity Effect in Sr2RuO4
Early studies of the impurity effects in Sr$_{2}$RuO$_{4}$ 
revealed several features reflecting the unconventional superconductivity: 
rapid suppression of $T_{\rm c}$ by native impurities and defects \cite{defect} and 
a large enhancement of the residual density of states in the superconducting state seen in  
specific heat measurements \cite{NishiZaki} and NMR measurements \cite{Ishida}. 
Throughout the above series of studies, 
control of the impurity concentration within the crystals was not easy, 
because the impurities 
were introduced accidentally during crystal growth. 
After considerable effort to optimize the growth conditions \cite{Mao-crystal}, 
we can now constantly obtain high quality crystals 
with minimal accidental contamination and $T_{\rm c}$ $>$ 1.4~K. 
%
%Substitution Effect
%
%Thus, 
%we have studied the substitution effect by Ru$^{4+}$ 
%in order to understand the impurity effect in more detail, 
%analogous to the nonmagnetic Zn$^{2+}$ (3$d^{10}$) and 
%magnetic Ni$^{2+}$ (3$d^{8}$) substitution in the high-$T_{\rm c}$ cuprates. 
%
This had allowed us to embark on a systematic study of the effects of controlled substitution of 
Ru$^{4+}$ with nonmagnetic Ti$^{4+}$ and magnetic Ir$^{4+}$ ions. 
The effect of impurity substitution into correlated electron systems can be subtle, 
with even nonmagnetic ions introducing magnetic effects. 
Before describing the scattering effects of very low concentrations of Ti$^{4+}$ and Ir$^{4+}$ 
on the superconductivity, 
it is therefore useful to review the effects of higher level doping on the magnetic properties. 
Recently, 
we reported that the substitution of 
\textit{nonmagnetic} impurity Ti$^{4+}$ (3$d^{0}$) 
in Sr$_{2}$RuO$_{4}$ induces a local moment with the effective moment 
$p_{\rm eff}$ $\sim$ 0.5 $\mu_{\rm B}$/Ti~\cite{Minakata}. 
The induced moment has Ising anisotropy 
with an easy axis along the $c$ direction. 
Furthermore, 
magnetic ordering with glassy behavior appears 
for $x$(Ti) $\geq$ 2.5\% in Sr$_{2}$Ru$_{1-x}$Ti$_{x}$O$_{4}$ 
while keeping metallic conduction along the in-plane direction. 
When $x$(Ti) is further increased to 9\%, 
$elastic$ neutron scattering measurements detect an incommensurate Bragg peak \cite{Braden} 
whose wave vector $\textit{\textbf{Q}}_{\rm ic}$ $\sim$ (2$\pi$/3, 2$\pi$/3, 0) 
is close to the position of the $inelastic$ neutron scattering peak seen in pure Sr$_{2}$RuO$_{4}$ \cite{Sidis}. 
%
%By the $elastic$ neutron scattering measurement, 
%incommensurate Bragg peak with the wave vector 
%$\textit{\textbf{Q}}_{\rm ic}$ $\sim$ (2$\pi$/3, 2$\pi$/3, 0) 
%is also observed for $x$(Ti) = 9\% \cite{Braden}, 
%where the $\textit{\textbf{Q}}_{\rm ic}$ is close to the position of the $inelastic$ 
%neutron scattering peak 
%seen in pure Sr$_{2}$RuO$_{4}$ \cite{Sidis}. 
%
In the vicinity of the magnetic ordering with $x$ $\geq$ 2.5\%, 
deviation from the Fermi-liquid behavior seen in the pure Sr$_{2}$RuO$_{4}$ is observed 
with the resistivity and the specific heat data showing linear-temperature dependence 
and logarithmic temperature dependence, respectively \cite{Kikugawa}. 
These results indicate that the two-dimensional antiferromagnetic spin fluctuations 
at $\textit{\textbf{Q}}_{\rm ic}$ arising from the nesting mainly in the $\beta$ band 
becomes a static spin density wave (SDW) by $nonmagnetic$ Ti substitution. 
On the other hand, 
the system Sr$_{2}$Ru$_{1-x}$Ir$_{x}$O$_{4}$ in which the substitution is $magnetic$ Ir$^{4+}$ (5$d^{5}$ in the low spin configuration) shows weak ferromagnetism at $x$(Ir) $\geq$ 30\% 
occurring concomitantly with metal-insulator transition~\cite{Cava} and 
the end-member material Sr$_{2}$IrO$_{4}$ is a Mott insulator with canted antiferromagnetic ordering~\cite{Cao}. 
% 
%where Ir$^{4+}$ (5$d^{5}$) would be considered as magnetic impurity with $S$ = 1/2 
%in the low spin configuration. 
%
Thus, 
substitution of high levels of Ti$^{4+}$ and Ir$^{4+}$ impurities in Sr$_{2}$RuO$_{4}$ 
leads to different magnetic ground states, 
presumably reflecting the different magnetic character of the isolated ions. 
%
%However, 
%there has not been a systematic report of similarity and difference of 
%the $nonmagnetic$ and the $magnetic$ impurity effects on the transport properties 
%related to the pair breaking. 

%Purpose
%
In this letter, 
we show that both nonmagnetic Ti$^{4+}$ and magnetic Ir$^{4+}$ impurities 
suppress the superconductivity of Sr$_{2}$RuO$_{4}$ completely by a concentration of $\sim$ 0.15\%. 
Also, 
both impurities act as strong potential scatterers with the maximum phase shift $\delta_{0}$ $\sim$ $\pi$/2 
(unitarity limit), 
as seen in high-$T_{\rm c}$ cuprates with $nonmagnetic$ Zn$^{2+}$ (3$d^{10}$) impurity. 
A series of single crystals of Sr$_{2}$Ru$_{1-x}$Ti$_{x}$O$_{4}$ and 
Sr$_{2}$Ru$_{1-x}$Ir$_{x}$O$_{4}$ with $x$ up to 3\%  were 
grown by a floating-zone method with an infrared image furnace (NEC Machinery, model SC-E15HD). 
The detailed procedure of the crystal growth is described elsewhere \cite{Mao-crystal,Minakata}. 
We only note here that an excess of 0.15 molar Ru was added for each 2 molar Sr for the crystal growth. 
The Ti and Ir concentrations in grown crystals were analyzed by electron-probe microanalysis (EPMA). 
The Ti is well substituted for Ru as reported by Minakata and Maeno~\cite{Minakata}. 
On the other hand, 
we found that 
the Ir as well as Ru was heavily evaporated during crystal growth 
at the high temperature of $\sim$2200~$^{\circ}$C, 
so that excess amounts had to be added for the crystal growth: 
the analyzed Ir concentration $x_{\rm a}$(Ir) is roughly connected with 
the nominal concentration $x_{\rm n}$(Ir) by $x_{\rm a}$ $\sim$ 0.25$x_{\rm n}$ for $x_{\rm n}$ $\leq$ 9\%.  
We note that the tetragonal crystal symmetry \cite{Minakata,Cava} for 
both Sr$_{2}$Ru$_{1-x}$Ti$_{x}$O$_{4}$ and 
Sr$_{2}$Ru$_{1-x}$Ir$_{x}$O$_{4}$ with $x$ up to 3\% was confirmed at room temperature. 
The induced moment by magnetic impurities is $isotropic$ in sharp contrast to the result of Ti substituted system 
and estimated as 0.7 $\mu_{\rm B}$/Ir from susceptibility measurements. 
This value is much smaller than the previous report using polycrystals, 
$\sim$2 $\mu_{\rm B}$/Ir at $x$(Ir) $\sim$ 5\% \cite{Cava}. 

\begin{figure}
\begin{center}
	\epsfxsize=7.5cm
	\epsfbox{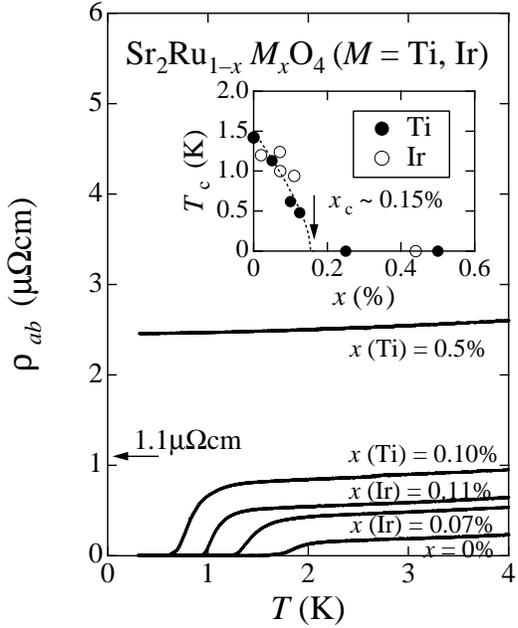}
\end{center}
\caption{
Temperature~dependence of the in-plane resistivities $\rho_{ab}$ 
in Sr$_{2}$Ru$_{1-x}$Ti$_{x}$O$_{4}$ and Sr$_{2}$Ru$_{1-x}$Ir$_{x}$O$_{4}$. 
Inset: Superconducting transition temperature $T_{\rm c}$ 
as a function of the impurity concentration $x$. 
The broken line shows the best fitting by Abrikosov-Gor'kov pair-breaking function. 
}
\label{Rho_ab_vs_T}
\end{figure}

For the resistivity measurements, 
the crystals were cut into rectangles with a typical size of 3.5 $\times$ 0.4 $\times$ 0.05 mm$^{3}$.  
The shortest dimension was along the $c$ axis. % perpendicular to the RuO$_{2}$ plane. 
Silver paste (Dupont, 6838) was used for attaching electrodes 
and cured at 500~$^{\circ}$C for 5 minutes;
the contact resistances were below 0.4~$\Omega$. 
The in-plane resistivity $\rho_{ab}$ measurements were performed 
by a standard four-probe dc method 
between 4.2 and 300~K and by a low frequency ac method between 0.3 and 5~K. 
Before measuring the $\rho_{ab}$ of Sr$_{2}$Ru$_{1-x}$Ti$_{x}$O$_{4}$ and 
Sr$_{2}$Ru$_{1-x}$Ir$_{x}$O$_{4}$, 
we examined the absolute value of the $\rho_{ab}$ 
for Sr$_{2}$RuO$_{4}$ crystals without Ti and Ir substitutions but with various $T_{\rm c}$ (1.42~K, 1.24~K and 
less than 0.3~K) in order to remove the uncertainty due to the size error and the inhomogeneous current path. 
The resistivity of these crystals were 121 $\pm$ 2 $\mu\Omega$cm at 300~K. 
Moreover, the residual resistivity $\rho_{ab0}$ %for the $T_{\rm c}$ = 1.42~K, 1.24~K and $<$~0.3~K 
were 0.15, 0.40 and 1.6 $\mu\Omega$cm, respectively, 
in the order of decreasing $T_{\rm c}$. 
Here,  
the $\rho_{ab0}$ was defined by the extrapolation of the the low temperature resistivity to $T = 0$. 
The $T_{\rm c}$ vs. $\rho_{ab0}$ agreed well with the previous data \cite{defect}. 
Figure \ref{Rho_ab_vs_T} shows the temperature dependence of $\rho_{ab}$ 
in Sr$_{2}$Ru$_{1-x}$Ti$_{x}$O$_{4}$ and Sr$_{2}$Ru$_{1-x}$Ir$_{x}$O$_{4}$ 
with a small amount of $x$. 
The $T_{\rm c}$ is rapidly and systematically suppressed in both cases.  
The result reflects the high sensitivity to translational symmetry breaking, 
characteristic of unconventional superconductivity. 
The inset shows the dependence of $T_{\rm c}$ on the impurity concentration $x$.  
We can see an almost universal suppression of $T_{\rm c}$, 
irrespective of the kind of impurity. 
The broken line shows the universal Abrikosov-Gor'kov pair-breaking function, 
where the formulation is generalized to the case of nonmagnetic and magnetic impurities 
in an unconventional superconductor~\cite{Theory}. 
Based on this model, $T_{\rm c}$($x$) satisfies 
\[ \ln\left(\frac{T_{\rm c}}{T_{\rm c0}} \right) = 
\mathit{\Psi}\left(\frac{\rm 1}{\rm 2} \right) - 
\mathit{\Psi}\left(\frac{\rm 1}{\rm 2} + \frac{\hbar\mathit{\Gamma}}{{\rm 2}\pi k_{\rm B}T_{\rm c}}\right). \] 
Here $\mathit{\Psi}$ is the digamma function, 
$\hbar$ the Dirac constant and the scattering rate 
$\displaystyle \mathit{\Gamma} = \frac{\rm 1}{{\rm 2}\tau} = \frac{{\rm 2}x}{{\pi}{\hbar}N_{\rm 0}} {\rm \sin^{2}\delta_{0}} + AS(S +\rm 1)$; 
the first and second terms in $\mathit{\Gamma}$ represent the potential and magnetic spin-flip scattering contributions, respectively, 
where $N_{0}$ is the density of states in the normal state. 
From our best fitting by fixing $T_{c0}$ as 1.5~K, 
the initial rate d$T_{\rm c}$/d$x$ $\sim$ $-7.5$ K/$x$(\%) is obtained 
for both Ti and Ir substitutions. 
The critical concentration $x_{\rm c}$ for disappearance of the superconductivity is estimated 
as $x_{\rm c}$ $\sim$ 0.15\%. 
%

\begin{figure}
\begin{center}
	\epsfxsize=7.5cm
	\epsfbox{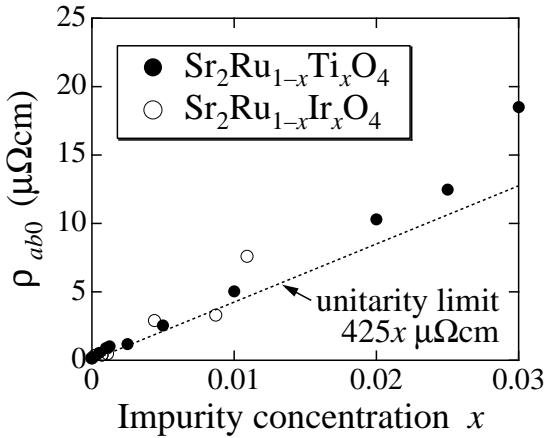}
\end{center}
\caption{
Residual resistivity $\rho_{ab0}$ as a function of $x$ for 
Sr$_{2}$Ru$_{1-x}$Ti$_{x}$O$_{4}$ and Sr$_{2}$Ru$_{1-x}$Ir$_{x}$O$_{4}$. 
The broken line represents the unitarity scattering with the phase shift $\delta_{0}$ = $\pi$/2. 
}
\label{UnitarityLimit}
\end{figure}

%
%In addition to the pair-breaking, 
%a rapid enhancement of the residual resistivity $\rho_{ab0}$ 
%was observed with both kinds of impurities. 
%
By measuring the residual resistivity $\rho_{ab0}$ (Fig. 1), 
we can see a universal trend that 
the superconductivity of Sr$_{2}$RuO$_{4}$ is completely suppressed 
at the critical resistivity of $\rho_{ab0}$ $\sim$ 1.1~$\mu$$\Omega$cm for both impurities, 
as reported from previous studies with native impurities and defects~\cite{defect}. 
The critical value $\rho_{ab0}$ is again in good agreement 
with the mean free path $l_{ab}$ falling below the superconducting coherence length 
$\xi_{ab}$ $\sim$ 900\AA when superconductivity is destroyed. 
The fact that Ti$^{4+}$ and Ir$^{4+}$ suppress $T_{\rm c}$ in the same way 
in spite of their very different magnetic characters suggests that 
the magnetic contribution to pair breaking is negligible, 
and potential scattering dominates. 
Although it could also be explained in other ways, 
this observation is qualitatively consistent with the existence of a spin-triplet state. 
Magnetic impurities break singlet pairs essentially 
because of exchange splitting of the single particle state; 
equal spin paired triplet states would not be subject to such an effect. 
Similar observations of negligible magnetic pair breaking have also been reported 
in UPt$_{3}$ \cite{UPt3}. 
In Fig.~\ref{UnitarityLimit}, 
the impurity concentration dependence of $\rho_{ab0}$ is displayed for 
Sr$_{2}$Ru$_{1-x}$Ti$_{x}$O$_{4}$ and Sr$_{2}$Ru$_{1-x}$Ir$_{x}$O$_{4}$. 
The enhancement of $\rho_{ab0}$ shows the same behavior for both impurities 
with a slope d$\rho_{ab0}$/d$x$ $\sim$ 500 $\mu\Omega$cm/$x$. 
For potential scattering, 
the residual resistivity in a two dimensional system is given as 
\[\rho_{ab0} = \frac{4\hbar}{e^{2}}\frac{x}{\sum_{i}^{\alpha,\beta,\gamma} n_{i}}\sin^{2}\delta_{0}, \] 
where, 
$n_{i}$ is the carrier concentration for each Fermi surface ($\alpha, \beta, \gamma$) \cite{Andy-dHvA}.
Using the relation $n_{i} = k_{\rm F \it i}^{2}/2\pi d$, 
where $k_{\rm F \it i}$ is each Fermi wave number in cylindrical Fermi surface approximation \cite{Andy-dHvA} and $d$ the interlayer distance, 
we can obtain d$\rho_{ab0}$/d$x$ = 425 $\mu\Omega$cm/$x$  
in Sr$_{2}$RuO$_{4}$, 
drawn as a broken line in Fig. \ref{UnitarityLimit}. 
Here we have assumed the unitarity limit, namely with the maximum phase shift $\delta_{0} = \pi/2$. 
The estimated value is 
in good agreement with the experimental results. 
Also, 
by assuming $only$ potential scattering contribution with $\delta_{0} = \pi/2$, 
we estimated 
$\displaystyle \frac{{\rm d}T_{\rm c}}{{\rm d}x} =  -\frac{\pi\hbar\mathit{\Gamma}}{2k_{\rm B}}\frac{1}{x}$
$\sim$ $-10$~K/$x$(\%) and the critical concentration for disappearance of the superconductivity 
$\displaystyle x_{\rm c} \sim \frac{\pi^{2}k_{\rm B}T_{\rm c0}N_{\rm 0}}{4\gamma}\frac{1}{\sin^{2}\delta_{0}}$
$\sim$ 0.1\%. 
Here, $\gamma$ is the Euler constant. 
These values  are  
consistent with the experimental values $\sim$ $-7.5$~K/$x$\% and $\sim$ 0.15\%, respectively. 
These results again suggest that both nonmagnetic and magnetic impurities 
act mainly as strong potential scatterers in Sr$_{2}$RuO$_{4}$. 
This unitarity scattering in Sr$_{2}$RuO$_{4}$ is similar to 
the substitution effect of nonmagnetic (Zn$^{2+}$) impurity  
in high-$T_{\rm c}$ cuprates \cite{high-Tc}. 
%
%Also, spin-flip scattering, 
%occurring in a metallic system with magnetic impurities \cite{Yoshida}, 
%seem to be negligible for magnetic Ir$^{4+}$-substituted Sr$_{2}$RuO$_{4}$. 
%
%Such absence of additional magnetic scattering is also discussed in other superconductor 
%UPt$_{3}$ with odd parity \cite{UPt3}. 
%

%
Table \ref{Table} summarizes the nonmagnetic and magnetic substitution effects in 
Sr$_{2}$RuO$_{4}$, 
in comparison with those in the high-$T_{\rm c}$ cuprates. 
For both impurities, 
further substitution leads to different magnetic ground state, 
namely spin glass behavior coexisting with incommensurate magnetic order \cite{Minakata,Braden} and 
weak ferromagnetism \cite{Cava,Cao}
for nonmagnetic and magnetic impurities, respectively. 
At very low doping levels, however, 
we have shown here that both impurities have very similar effects on transport and magnetic properties. 
%
%Although, 
%we can see similar effects on transport and magnetic properties for both impurities with small amount 
%in Sr$_{2}$RuO$_{4}$.  
%
In order to clarify the similarity and the difference in more detail, 
it is very important to investigate how the spin fluctuation at $\textit{\textbf{Q}}_{\rm ic}$ 
is modified by the magnetic impurity, as has been done for the nonmagnetic impurity~\cite{Braden}. 

In summary, 
we report systematic comparison of the substitution effects of $nonmagnetic$ and $magnetic$ impurities 
in the spin-triplet superconductor Sr$_{2}$RuO$_{4}$. 
Irrespective of leading to different magnetic ordering at high levels of substitution,  
we found universal behavior for both impurities in the suppression of $T_{\rm c}$ and 
the enhancement of $\rho_{ab0}$, in accordance with the strong potential scattering 
with $\delta_{0}$ = $\pi$/2. 
Our result suggests that both nonmagnetic and magnetic impurities in Sr$_{2}$RuO$_{4}$ 
break pairs due to strong potential scattering, 
and that magnetic scattering does not play an important role.  

The authors thank 
H. Fukazawa, K. Deguchi and M. Yoshioka, P.D.A. Mann and D. Herd for useful discussions and technical support. 
They also thank T. Nomura and M. Sigrist for useful discussions, 
Y. Shibata, J. Hori and T. Fujita for EPMA measurements at Hiroshima University. 
This work was mainly supported  by Core Research for Evolutional Science and Technology (CREST) of 
Japan Science and Technology Corporation, 
and by the Grant-in-Aid for Scientific Research on Priority Area 
"Novel Quantum Phenomena in Transition Metal Oxides" from the Ministry of Education, Culture, 
Sports, Science and Technology.  
%

%References

%\clearpage

%Figures%%%%%%%%%%%%%%%%%%%

%%%%%%%%%%%%%%%%%%%%%%%%%%%%%%%%%%%%%%%%
\begin{table}
\caption{Substitution effects of nonmagnetic and magnetic impurity in 
(a) Sr$_{2}$RuO$_{4}$, (b) underdoped and (c) overdoped cuprates.}
\label{Table}

\vspace{3mm}
(a) Sr$_{2}$RuO$_{4}$
\begin{tabular}{@{\hspace{\tabcolsep}\extracolsep{\fill}}ccc} \hline
& Sr$_{2}$Ru$_{1-x}$Ti$_{x}$O$_{4}$  & Sr$_{2}$Ru$_{1-x}$Ir$_{x}$O$_{4}$ \\ \hline
phase shift ($\delta_{0}$)	& $\pi$/2 & $\pi$/2\\
$p_{\rm eff}$	& 0.5~$\mu_{\rm B}$/Ti~\cite{Minakata} & 0.7~$\mu_{\rm B}$/Ir \\
magnetic order	& $x$(Ti) $\geq$ 2.5\%~\cite{Minakata}	& $x$(Ir) $\geq$ 30\%~\cite{Cava} \\
	& SDW and spin glass & weak ferromagnetism~\cite{Cava} \\\hline
\end{tabular}

\vspace{3mm}
(b) high-$T_{\rm c}$ cuprates (underdoped)
\begin{tabular}{@{\hspace{\tabcolsep}\extracolsep{\fill}}ccc} \hline
& nonmagnetic (Zn$^{2+}$)  & magnetic (Ni$^{2+}$) \\ \hline
phase shift ($\delta_{0}$)	& $\pi$/2~\cite{high-Tc} & 0.36$\pi$~\cite{high-Tc}\\
$p_{\rm eff}$	& 1~$\mu_{\rm B}$/Zn~\cite{Mendels} & 1.6~$\mu_{\rm B}$/Ni~\cite{Mendels} \\
magnetic order	& $-$	& $-$ \\\hline
\end{tabular}

\vspace{3mm}
(c) high-$T_{\rm c}$ cuprates (overdoped)
\begin{tabular}{@{\hspace{\tabcolsep}\extracolsep{\fill}}ccc} \hline
& nonmagnetic (Zn$^{2+}$)  & magnetic (Ni$^{2+}$) \\ \hline
phase shift ($\delta_{0}$)	& $\pi$/2~\cite{high-Tc} & 0.32 - 0.36$\pi$~\cite{Uchida}\\
$p_{\rm eff}$	& 0.4~$\mu_{\rm B}$/Zn~\cite{Mendels} & 1.2~$\mu_{\rm B}$/Ni~\cite{Mendels} \\
magnetic order	& $-$	& $-$ \\\hline
\end{tabular}

\end{table}

%%%%%%%%%%%%%%%%%%%%%%%%%%%%%%%%%%%%%%%%%%%%%%%%%%%%%


\begin{thebibliography}{99}

\bibitem{Mathur}
For instance, N.D. Mathur, F.M. Grosche, S.R. Julian, I.R. Walker, D.M. Freye, R.K.W. Haselwimmer and G.G. Lonzarich: Nature \textbf{394} (1998) 39. 

\bibitem{unconventionalSC}
For instance, Y. Kitaoka, K. Ishida and K. Asayama: J. Phys. Soc. Jpn. \textbf{63} (1994) 2052. 

\bibitem{UPt3}
Y. Dalichaouch, M.C. de Andrade, D.A. Gajewski, R. Chau, P. Visani and M.B. Maple: 
Phys. Rev. Lett. \textbf{75} (1995) 3938; 
H.G.M. Duijn, N.H. van Dijk, A. de Visser and J.J.M. Franse: Physica B \textbf{223\&224} (1996) 44. 

\bibitem{HFSC}
For instance, 
F. Kromor, M. Lang, N. Oeschler, P. Hinze, C. Langhammer and F. Steglich: 
Phys. Rev. B \textbf{62} (2000) 12477, and references therein; 
B. Arfi, H. Bahlouli, C.J. Pethick and D. Pines: Phys. Rev. Lett. \textbf{60} (1988) 2206. 

\bibitem{high-Tc}
For instance, 
Y.Fukuzumi, K. Mizuhashi, K. Takenaka and S. Uchida: Phys. Rev. Lett. \textbf{76} (1996) 684; 
S.H. Pan, E.W. Hudson, K.M. Lang, H. Eisaki, S. Uchida and J.C. Davis: Nature \textbf{403} (2000) 746; 
E.W. Hudson, K.M. Lang, V. Madhavan, S.H. Pan, H. Eisaki, S. Uchida and J.C. Davis: Nature \textbf{411} (2001) 920. 

\bibitem{Nature} 
Y. Maeno, H. Hashimoto, K. Yoshida, A. Nishizaki, T. Fujita, J.G. Bednorz and F. Lichtenberg: 
Nature \textbf{372} (1994) 532. 

\bibitem{physicstoday}
Y. Maeno, T.M. Rice and M. Sigrist: Physics Today \textbf{54} (2001) 42 and references therein. 

\bibitem{Maeno}
Y. Maeno, K. Yoshida, H. Hashimoto, S. Nishizaki, S. Ikeda, M. Nohara, T. Fujita, A.P. Mackenzie, N.E. Hussy, J.G. Bednorz and F. Lichtenberg: J. Phys. Soc. Jpn. \textbf{66} (1997) 1405. 

\bibitem{Andy-dHvA}
A.P. Mackenzie, S.R. Julian, A.J. Diver, G.J. McMullan, M.P. Ray, G.G. Lonzarich, Y. Maeno, S. Nishizaki and T. Fujita: Phys. Rev. Lett. \textbf{76} (1996) 3786. 

\bibitem{Band}
T. Oguchi: Phys. Rev. B \textbf{51} (1995) R1385; 
D.J. Singh: Phys. Rev. B \textbf{52} (1995) 1358. 

\bibitem{defect}
A.P. Mackenzie, R.K.W. Haselwimmer, A.W. Tyler, G.G. Lonzarich, Y. Mori, S. Nishizaki and Y. Maeno: 
Phys. Rev. Lett. \textbf{80} (1998) 161; 
Z.Q. Mao, Y. Mori and Y. Maeno: Phys. Rev. B. \textbf{60} (1999) 610.

\bibitem{NishiZaki}
S. NishiZaki, Y. Maeno and Z.Q. Mao: J. Low Temp. Phys. \textbf{117} (1999) 1581. 

\bibitem{Ishida}
K. Ishida, H. Mukuda, Y. Kitaoka, Z.Q. Mao, H. Fukazawa and Y. Maeno: 
Phys. Rev. Lett. \textbf{84} (2000) 5387. 

\bibitem{Mao-crystal}
Z.Q. Mao, Y. Maeno and H. Fukazawa: Mat. Res. Bull. \textbf{35} (2000) 1813. 

\bibitem{Minakata}
M. Minakata and Y. Maeno: Phys. Rev. B. \textbf{63} (2001) R180504.

\bibitem{Braden}
M. Braden, O. Friedt, Y. Sidis, P. Bourges, M. Minakata and Y. Maeno: 
Phys. Rev. Lett. \textbf{88} (2002) 197002.

\bibitem{Sidis}
Y. Sidis, M. Braden, P. Bourges, B. Hennion, S. NishiZaki, Y. Maeno and Y. Mori: 
Phys. Rev. Lett. \textbf{83} (1999) 3320. 

\bibitem{Kikugawa}
N. Kikugawa and Y. Maeno: 
Phys. Rev. Lett. \textbf{89} (2002) 117001. 

\bibitem{Cava}
R.J. Cava, B. Batlogg, K. Kiyono and H. Takagi: Phys. Rev. B. \textbf{49} (1994) 11890.

\bibitem{Cao}
G. Cao, J. Bolivar, S. McCall, J.E. Crow and R.P. Guertin: Phys. Rev. B. \textbf{57} (1998) 11039; 
M.K. Crawford, M.A. Subramanian, R.L. Harlow, J.A. Fernandez-Baca, Z.R. Wang and D.C. Johnston: 
Phys. Rev. B. \textbf{50} (1994) 9419.

\bibitem{Theory}
For instance, R.J. Radtke, K Levin, H.-B. Sch$\ddot{\rm u}$ttler and M.R. Norman: Phys. Rev. B. \textbf{48} (1993) 653.

%\bibitem{Adachi}
%H. Adachi and R. Ikeda: J. Phys. Soc. Jpn. \textbf{70} (2001) 2848. 

%\bibitem{Yoshida}
%T. Kasuya: Prog. Theor. Phys. \textbf{22} (1959) 227. 

\bibitem{Mendels}
P. Mendels, J. Bobroff, G. Collin, H. Alloul, M. Gabay, J.F. Marucco, N. Blanchard and B. Grenier: 
Europhys. Lett. \textbf{46} (1999) 678. 

\bibitem{Uchida}
S. Uchida and K.M.  Kojima: private communication. 

\end{thebibliography}
\end{document}